\documentclass[bibyear]{aa} % if the references are not structured
\usepackage{graphicx}
\usepackage{amsmath}
\usepackage{float} % para usar [H]
\usepackage{caption}

\usepackage{txfonts}
\begin{document}

%%%%%%%%%%%%%%%%%%%%%%%%%%%%%%%%%%%%%%%%%%%%%%%%%%%%%%%%%%%%%%%%%%%%%%%%%%%%%%%%%%%%%%%%%%%%%%%%%%%%%%%%%%%%%%%%%%%%%%%%%%%%%%%%%%%%%%%%%%%%%%%%%%%%%%%%%%%%%%%%%%%%%%%%%%%%%%%%%%%%%%%%%%%%%%%%%%%%%%%%%%%%%%%%%%

\title{Power-2 limb-darkening coefficients for the $uvby$, $UBVRIJHK$, SDSS $ugriz$, \textit{Gaia}, \textit{Kepler,} TESS, and CHEOPS photometric systems
        \thanks{Tables 1-12 are only available at the CDS via anonymous ftp to cdsarc.u-strasbg.fr (130.79.128.5) or via http://cdsarc.u-strasbg.fr/viz-bin/cat/J/A+A/XXX/XXX }}

\subtitle{II. PHOENIX spherically symmetric stellar atmosphere models}

\titlerunning {Limb-darkening coefficients for the power-2 law}

\author{A. Claret~\inst{1,2}  \and J. Southworth \inst{3}}

%\offprints{A. Claret, e-mail:claret@iaa.es}

\institute{Instituto de Astrof\'{\i}sica de Andaluc\'{\i}a, CSIC, Apartado 3004, 18080 Granada, Spain
      \and Dept.\ F\'{\i}sica Te\'{o}rica y del Cosmos, Universidad de Granada, Campus de Fuentenueva s/n,  10871, Granada, Spain
      \and Astrophysics Group, Keele University, Staffordshire, ST5 5BG, UK
}

\date{Received; accepted; }

\abstract
{The phenomenon of limb-darkening is relevant to many topics in astrophysics, including the analysis of light curves of eclipsing binaries, optical interferometry, measurement of stellar diameters, line profiles of rotating stars, gravitational microlensing, and transits of extrasolar planets}
{Multiple parametric limb-darkening laws have been presented, and there are many available sources of theoretical limb-darkening coefficients (LDCs) calculated using stellar model atmospheres. The power-2 limb-darkening law allows a very good representation of theoretically predicted intensity profiles, but few LDCs are available for this law from spherically symmetric model atmospheres. We therefore present such coefficients in this work.}
{We computed LDCs for the space missions \textit{Gaia}, \textit{Kepler}, TESS, and CHEOPS and for the passbands $uvby$, $UBVRIJHK$, and SDSS $ugriz$, using the \textsc{phoenix-cond} spherical models. We adopted two methods to characterise the truncation point, which sets the limb of the star: the first (M1) uses the point where the derivative d$I(r)$/d$r$ is at its maximum ---where I(r) is the specific intensity as a function of the normalised radius r ---  corresponding to $\mu_{\rm cri}$, and the second (M2) uses the midpoint between the point $\mu_{\rm cri}$ and the point located at $\mu_{\rm cri-1}$.
The LDCs were computed adopting the Levenberg-Marquardt least-squares minimisation method, with a resolution of 900 equally spaced $\mu$ points, and covering 823 model atmospheres for a solar metallicity, effective temperatures of 2300 to 12000\,K, $\log g$ values from 0.0 to 6.0, and microturbulent velocities of 2\,km\,s$^{-1}$. As our previous calculations of LDCs using spherical models included only 100 $\mu$ points, we also updated the calculations for the four-parameter law for the passbands listed above, and compared them with those from the power-2 law.}
{Comparisons between the quality of the fits provided by the power-2 and four-parameter laws show that the latter presents a lower merit function, $\chi^2$, than the former for both cases (M1 and M2). This is important when choosing the best approach for a particular science goal.}
{}


\maketitle

%%%%%%%%%%%%%%%%%%%%%%%%%%%%%%%%%%%%%%%%%%%%%%%%%%%%%%%%%%%%%%%%%%%%%%%%%%%%%%%%%%%%%%%%%%%%%%%%%%%%%%%%%%%%%%%%%%%%%%%%%%%%%%%%%%%%%%%%%%%%%%%%%%%%%%%%%%%%%%%%%%%%%%%%%%%%%%%%%%%%%%%%%%%%%%%%%%%%%%%%%%%%%%%%%%

\section{Introduction}

Stars show a decrease in specific intensity from their centre to their limb. This is because opacity causes the sightlines of the observer to penetrate less far when they enter the photosphere at an angle. This effect, known as limb-darkening, occurs both for real stars and for theoretical stellar atmosphere models, and is important in any situation where a star is spatially resolved, such as the study of eclipsing binaries, transiting planetary systems, stellar interferometry, microlensing, and spectral line profiles.

The specific intensity profiles generated by model atmospheres with spherical symmetry differ from their plane-parallel equivalents mainly in the region near the limb, where the former shows a sharp drop while the latter presents a finite specific intensity. This important difference is due to the fact that in  the plane-parallel case, the curvature of the atmosphere is neglected. Neglecting curvature effects is justified in  stars with low atmospheric scale heights, but not for cool giants or supergiants. In plane-parallel model calculations, the medium is infinite, there is no natural boundary condition, and the specific intensity is not zero at the limb but depends on the local temperature in this region. However, in the case of models with spherical symmetry, the temperature tends to zero at small optical depths (see Eq.~7.182 in Mihalas 1978). A concise explanation for this crucial difference can be found in Mihalas (1978, pp.~246-247); see also Larson (1969).  In the theoretical-observational context of flux and polarisation calculations, Kostogryz et al.\ (2017) report the superiority of spherical models over plane-parallel models in some configurations, such as systems with grazing eclipses, transits with Earth-size planets, or for hotter planet host stars (effective temperatures $T_{\rm eff} > 6000$\,K).

In Paper~I (Claret \& Southworth 2022), we showed the superiority of the power-2 law (Hestroffer 1997) in terms of the quality of fits of limb-darkening coefficients (LDCs) to theoretical predictions over the other two-parameter laws in the case of plane-parallel model atmospheres. However there remain few sources of LDCs calculated for the power-2 law, especially using spherical model atmospheres. The exceptions are LDCs for DA, DB, and DBA white dwarfs using plane-parallel models (Claret et al. 2020), solar-type stars using 3D models (Maxted 2018), extensive tables using the {\sc atlas} plane-parallel models (Claret \& Southworth 2022), and LDCs for the CHEOPS space mission, where spherical models are adopted (Claret 2021).

The aim of this work is to provide users with LDCs for the power-2 law based on spherical model atmospheres, covering a wide range of $T_{\rm eff}$ values, surface gravities, and passbands. The passbands adopted here are Str\"omgren $uvby$, Johnson-Cousins $UBVRIJHK$, and SDSS $ugriz$ photometric systems, plus those for the space missions \textit{Gaia}, \textit{Kepler}, TESS, and CHEOPS. The results of the present study supersede those of Claret (2021) for the CHEOPS mission.

The structure of the paper is as follows. Section 2 is dedicated to the description of the spherical models (\textsc{phoenix-cond}) and the numerical details of the calculations of the LDCs. A comparison between the power-2 and four-parameter LDCs is presented in Section 3. Section 4 gives a comparison between the plane-parallel and spherical LDCs, and Section 4 summarises our findings.

%%%%%%%%%%%%%%%%%%%%%%%%%%%%%%%%%%%%%%%%%%%%%%%%%%%%%%%%%%%%%%%%%%%%%%%%%%%%%%%%%%%%%%%%%%%%%%%%%%%%%%%%%%%%%%%%%%%%%%%%%%%%%%%%%%%%%%%%%%%%%%%%%%%%%%%%%%%%%%%%%%%%%%%%%%%%%%%%%%%%%%%%%%%%%%%%%%%%%%%%%%%%%%%%%%

\section{The spherical atmosphere models and numerical methods}

In this paper, we use specific intensity profiles from the \textsc{phoenix-cond} spherical model atmospheres. The main characteristics of these models are described in Husser et al.\ (2013); see also Claret et al.\ (2012). We adopt a Levenberg-Marquardt least-squares minimisation method to compute the LDCs. The specific intensities for the $uvby$, $UBVRIJHK$, $ugriz$, \textit{Gaia}, \textit{Kepler}, TESS, and CHEOPS photometric systems were integrated using the following equation:
\begin{equation}
I_{a}(\mu) =  ({hc})^{-1} {\int_{\lambda_1}^{\lambda_2} { I(\lambda,\mu) \lambda S(\lambda)_{a} d\,\lambda}\over\int_{\lambda_1}^{\lambda_2} { S(\lambda)_{a}~ d\,\lambda}}
,\end{equation}
where $h$ is Planck's constant, $c$ is the speed of light in vacuum, $\lambda$ is the wavelength in \AA,\ and $\mu$ is given by $\cos(\gamma)$, where $\gamma$ is the angle between the line of sight and the outward surface normal. $I_{a}(\mu)$ is the specific intensity for a given passband $a$, $I(\lambda,\mu)$ is the monochromatic specific intensity, and $S_a(\lambda)$ is the response function. For the $uvby$, $UBVRIJHK,$ and $ugriz$ photometric systems, the response function includes the transmission of one airmass of the Earth's atmosphere. The passbands used were obtained from Spanish Virtual Observatory Filter Profile Service\footnote{\texttt{http://svo2.cab.inta-csic.es/theory/fps/}}.  For the specific case of Str\"omgren $uvby$, the response function was obtained from the Observatorio de Sierra Nevada, Granada, Spain (C.\ C\'ardenas, private communication). For $JHK,$ the corresponding $S(\lambda)$ were obtained from the Observatorio del Teide-IAC, Spain (Alonso et al.\ 1994).

The specific intensities generated by the spherical model atmospheres show a much more pronounced curvature near the limb than the plane-parallel models, making it much more difficult to obtain a good fit to them. For example, the four-parameter law is able to fit the full profiles well but only for some filters (see e.g. Fig.~1 in Claret et al. (2012). Some alternative methods have been proposed to better describe the specific intensity profiles for models with spherical symmetry, such as Claret \& Hauschildt (2003), where the concept of quasi-spherical models was introduced.

Later, Wittkowski et al.\ (2004) introduced a more elaborate method: instead of truncating the models at a certain value of  $\mu$, the truncation was defined by searching for the maximum of the derivative of the specific intensity with respect to $r$, where $r = \sqrt{(1-\mu^2)}$. This point corresponds to $\tau_R \approx 1.0$, where $\tau_R$ is the Rosseland mean optical depth.  Fortunately, the {\sc phoenix} models were computed with sufficient points in the drop-off region to enable us to accurately determine the corresponding derivatives. Wittkowski et al.\ (2004) gave two ways to determine this critical point. The first method (M1) uses the point where the derivative is maximum, which is known as the point $\mu_{\rm cri}$, while the second method (M2) uses the midpoint between the point $\mu_{\rm cri}$ and the point located at $\mu_{\rm cri-1}$ (G.\ Morello, private communication). The specific intensity using M2 at this average point is larger than that obtained by the M1 method. Therefore, it gives rescaled profiles that are more similar to the plane-parallel models than those provided by M1. However, M2 does not accurately represent the specific intensity at the critical point as defined by Wittkowski et al.\ (2004). The corresponding critical points for both methods are characterised by $\mu_{\rm cri,1}$ and  $\mu_{\rm cri,2}$. In the current work, we present calculations for both M1 and M2 for completeness.

The LDC calculations were performed adopting 823 {\sc phoenix} stellar model atmospheres for solar metallicity, surface gravities $\log g$ from 0.0 to 6.0, $T_{\rm eff}$ values from 2300\,K to 12000\,K, and microturbulent velocities $V_\xi = 2.0$\,km\,s$^{-1}$. We adopted 900 equally spaced-points in $\mu$, rather than the 78 original ones adopted in the {\sc phoenix} models. In our previous papers on LDCs from spherical models, we used only 100 $\mu$ points, and so we have updated our calculations using 900 points for the photometric systems studied in this work.

The power-2 LD law introduced by Hestroffer (1997) is
\begin{equation}
\frac{I(\mu)}{I(\mu=1)}=  1 - g(1 - \mu^h) \,\, ,
\end{equation}
where $g$ and $h$ are the corresponding LDCs. Given that the profiles generated by {\sc phoenix} are much more complicated and difficult to adjust than in the plane-parallel approximation, we also calculated LDCs using the four-parameter law (Claret 2000) which is

\begin{equation}
\frac{I(\mu)}{I(\mu=1)} = 1 - \sum_{k=1}^{4} {a_k} (1 - \mu^{\frac{k}{2}}) \,\, ,
\end{equation}

where $a_k$ are the four LDCs. The greater flexibility of the four-parameter law allows it to reproduce the specific intensity profiles to a higher precision than limb-darkening laws using only two LDCs.

We define a merit function, which measures the quality of the fit to a given set of $I(\mu)$ values as
\begin{equation}
{\chi^2} \equiv \sum_{i=1}^{N} \left( {y_i - Y_i}\right)^2 \,\, ,
\end{equation}
where $y_i$ is the model intensity at point $i$, $Y_i$ is the fitted function at the same point, and $N$ is the number of $\mu$ points.

%%%%%%%%%%%%%%%%%%%%%%%%%%%%%%%%%%%%%%%%%%%%%%%%%%%%%%%%%%%%%%%%%%%%%%%%%%%%%%%%%%%%%%%%%%%%%%%%%%%%%%%%%%%%%%%%%%%%%%%%%%%%%%%%%%%%%%%%%%%%%%%%%%%%%%%%%%%%%%%%%%%%%%%%%%%%%%%%%%%%%%%%%%%%%%%%%%%%%%%%%%%%%%%%%%

\section{Results}

An extensive comparison of the different limb-darkening laws was carried out in Paper~I. It was found that the power-2 law produced better fits to the specific intensities from plane-parallel model atmospheres than the other two-parameter laws (quadratic, logarithmic, and square-root). Given this result and the increased complexity of the specific intensities from spherical model atmospheres, it was not thought necessary to repeat that analysis here. However, we find it useful to investigate the relative success of the power-2 and four-parameter laws, as this information is helpful to users of the LDCs.

\begin{figure}
        \includegraphics[height=8.9cm,width=6cm,angle=-90]{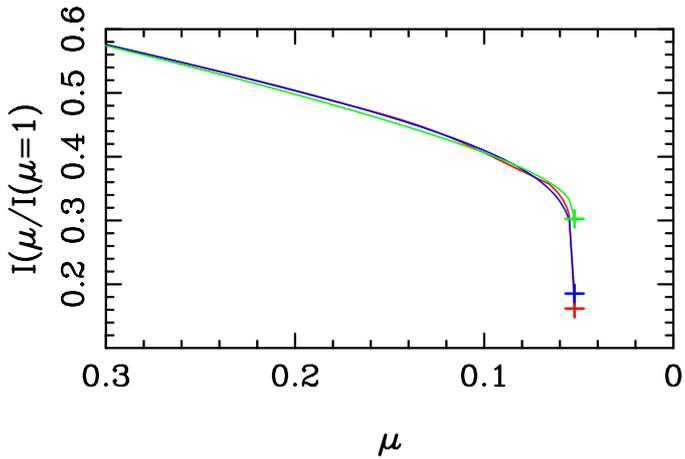}
        \caption{Angular distribution of the specific intensity for a model with
                $T_{\rm eff} = 4500$\,K, $\log g = 4.5$, [M/H] $=$ 0.0, and $V_\xi = 2$\,km\,s$^{-1}$ for the TESS passband.
                The red line represents the specific intensity distribution and the red cross indicates $\mu_{\rm cri,1}$.
                The green line denotes the fitting adopting the power-2 law, while the green cross indicates the fitting at $\mu_{\rm cri,1}$.
                The blue line denotes the four-parameter law approach and the blue cross indicates the fitting at $\mu_{\rm cri,1}$.
                This plot is for case M1, and the specific intensity profiles have not been re-scaled.}
\end{figure}

\begin{figure}
        \includegraphics[height=8.9cm,width=6cm,angle=-90]{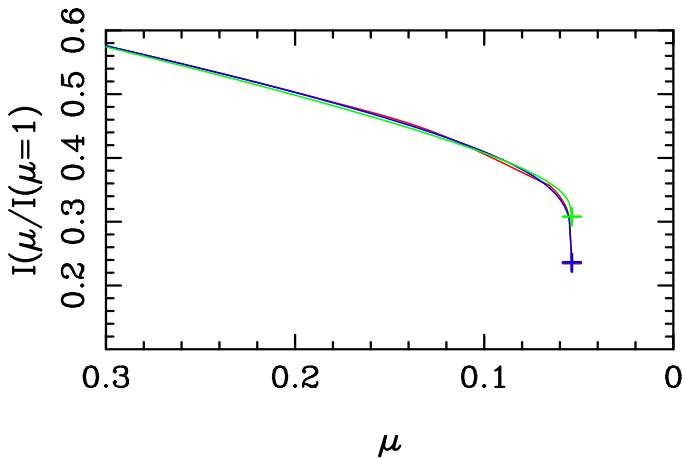}
        \caption{Same as in Fig.~1, except for case M2 and $\mu_{\rm cri,2}$ instead of case M1 and $\mu_{\rm cri,1}$.}
\end{figure}

\begin{figure}
        \includegraphics[height=8.9cm,width=6cm,angle=-90]{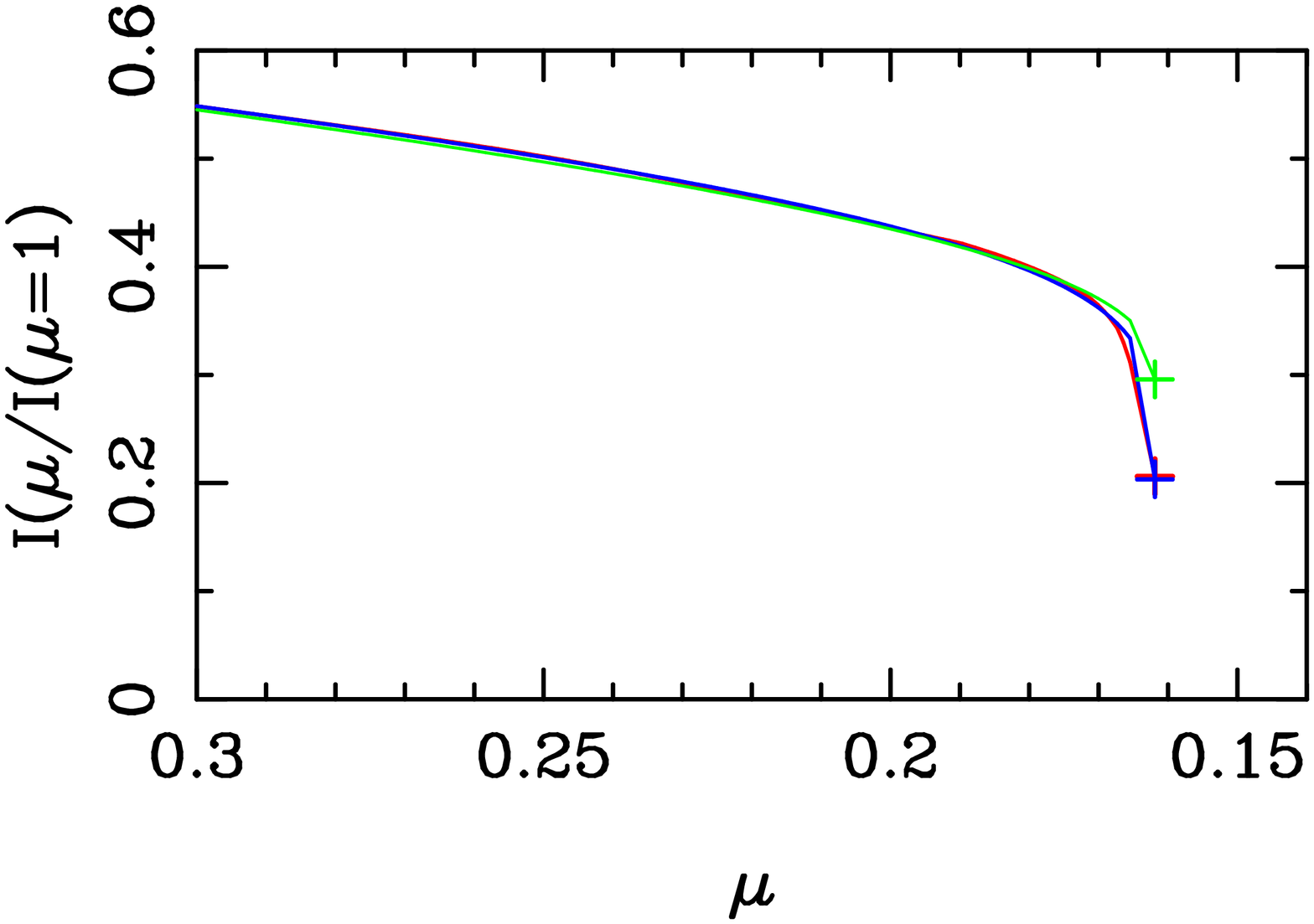}
                \caption{Same as in Fig.~1, except for $\log g = 2.5$ instead of $\log g = 4.5$.}
\end{figure}

\begin{figure}
        \includegraphics[height=8.9cm,width=6cm,angle=-90]{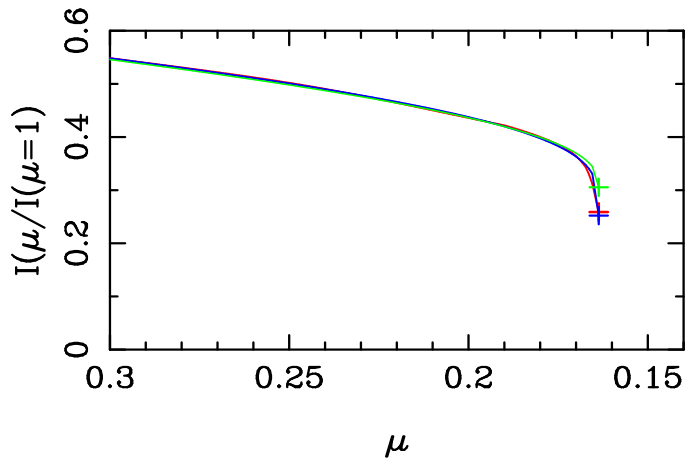}
                \caption{Same as in Fig.~2, except for $\log g = 2.5$ instead of $\log g = 4.5$.}
\end{figure}

The resulting $\chi^2$ values are generally small and are dominated by the difficulty of fitting the $\mu_{\rm cri,1}$ and $\mu_{\rm cri,2}$ points, which is most conspicuous for the power-2 laws in the M1 case. Figures\ 1--4 show the specific intensity profiles for the M1 and M2 cases using both the power-2 (Eq.~2) and four-parameter (Eq.~3) law. For the sake of clarity, only the regions near the limb are shown, because the fits for $\mu > \mu_{\rm cri}$ are almost perfect. For main-sequence stars (Figs.\ 1 and 2), and  in both the M1 and M2 cases, the four-parameter law (Eq. 3) provides better fits than those provided by the power-2 law (Eq. 2). The situation is the same but even clearer for the giant stars (Figs.\ 3 and 4). It is important to emphasise that one of the main differences between the plane-parallel models and the spherical ones is in the drop-off region, and so it is necessary to characterise this region well through the LDCs. The best test of the spherical models would be achieved using Eq.~3 and case M1.

The relative quality of the fits can be  computed using the following expression:
\begin{equation}
\alpha {\rm (passband)} = {{{\chi^2}\mbox{(power-2, case M1)}}\over{\chi^2 \mbox{(four-parameter, case M1)}}} \,\, .
\end{equation}
Figures\ 5--10 show the quantity $\log \alpha$ for the M1 case, a range of $T_{\rm eff}$ and $\log g$ values, and for the \textit{Gaia} $G_{\rm BP}$, $G$, $G_{\rm RP}$, \textit{Kepler}, TESS, and CHEOPS passbands. It can be seen that the $\chi^2$ provided by Eq.~3 is typically a factor of $8 \pm 2$ smaller than the $\chi^2$ given by Eq.~2, depending on the effective wavelength of the passband. We also see that there is a common maximum in all passbands centred on $\log T_{\rm eff} \approx 3.65$ ($\approx$4500\,K), where the four-parameter law performs much better than the power-2 law. A comparison of the effects for the SDSS $u$ and $z$ passbands (not shown) indicates that the ratio $\alpha$ is larger at shorter wavelengths, which is in agreement with the results for the \textit{Gaia} $G_{\rm BP}$ passband (Fig.~5). It is also apparent that the power-2 law has a better relative performance at higher $T_{\rm eff}$ values, but this is still  poorer than the four-parameter law.

\begin{figure}
        \includegraphics[height=8.9cm,width=6cm,angle=-90]{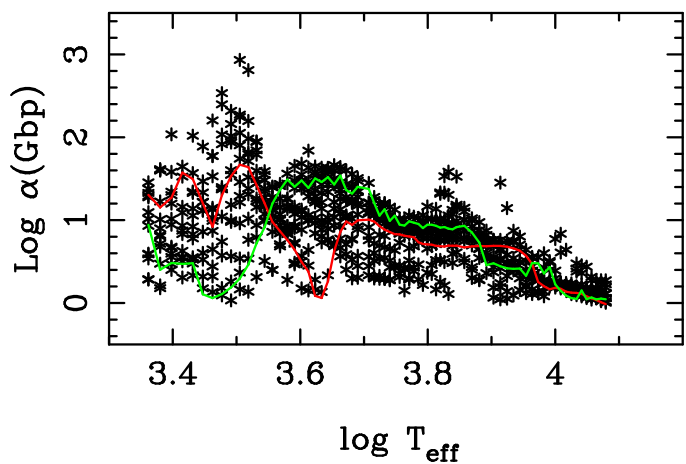}
        \caption{Comparison of the quality of fits provided for the \textit{Gaia} $G_{\rm BP}$ passband.
                The quantity $\alpha$ is the ratio between the $\chi^2$ for the power-2 law and the $\chi^2$ for the four-parameter law (both for case M1), for all 823 models.
                The red line traces the results for the most compact models ($\log g=6.0$) whilst the green line indicates the results for subgiant ones ($\log g=3.5$).}
\end{figure}

\begin{figure}
        \includegraphics[height=8.9cm,width=6cm,angle=-90]{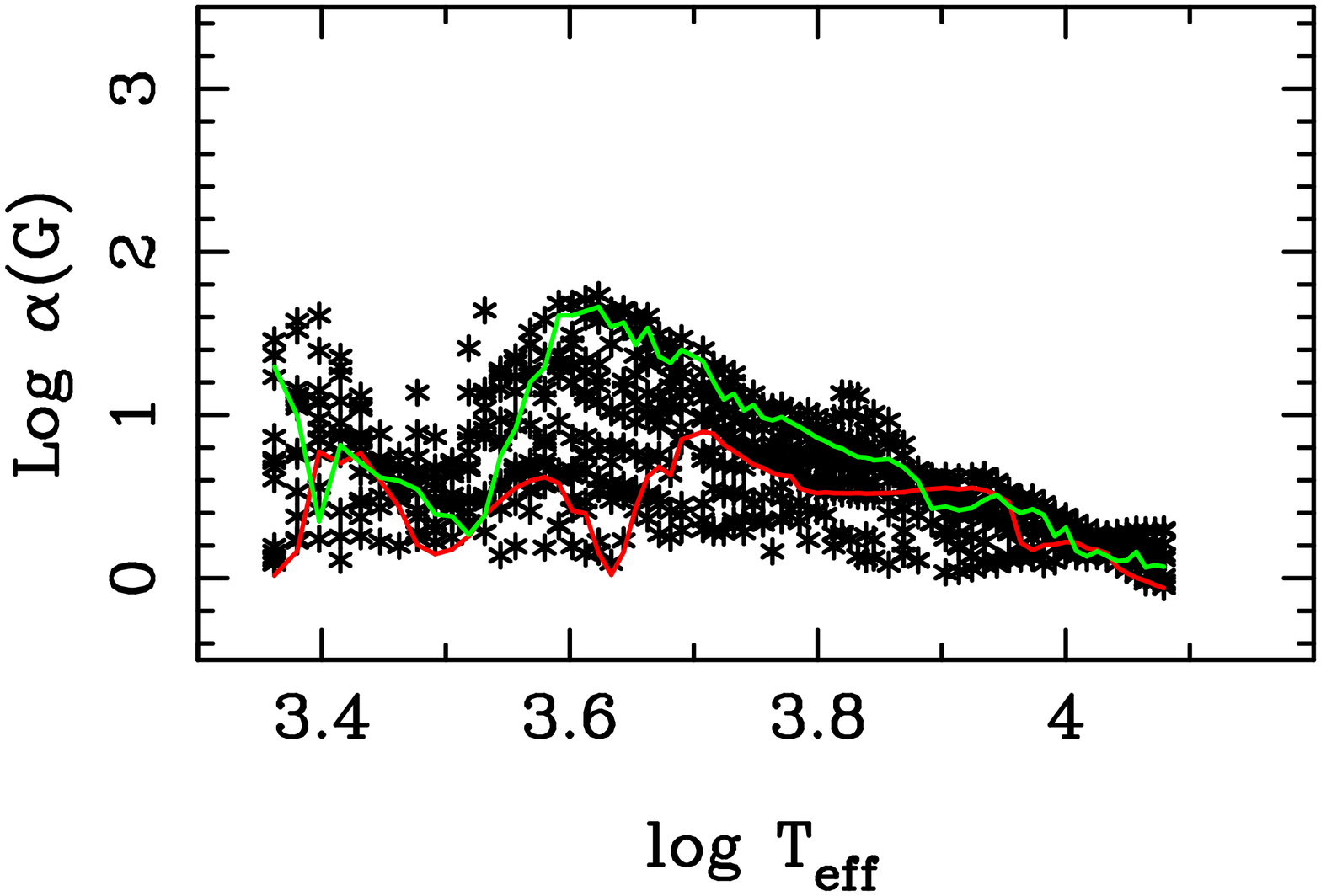}
        \caption{Same as in Fig.~5, but for the \textit{Gaia} $G$ passband.}
\end{figure}

\begin{figure}
        \includegraphics[height=8.9cm,width=6cm,angle=-90]{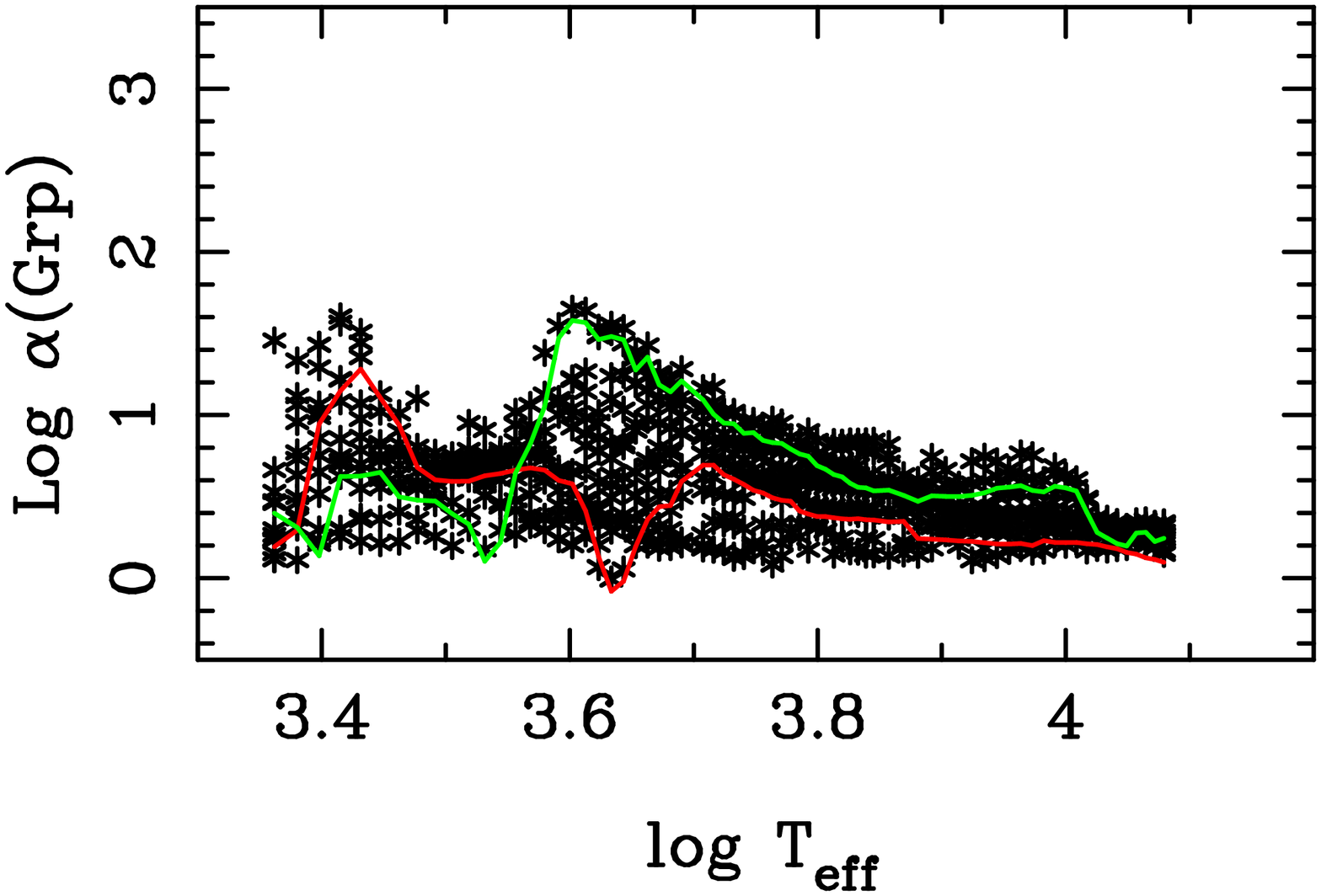}
        \caption{Same as in Fig.~5, but for the \textit{Gaia} $G_{\rm RP}$ passband.}
\end{figure}

\begin{figure}
        \includegraphics[height=8.9cm,width=6cm,angle=-90]{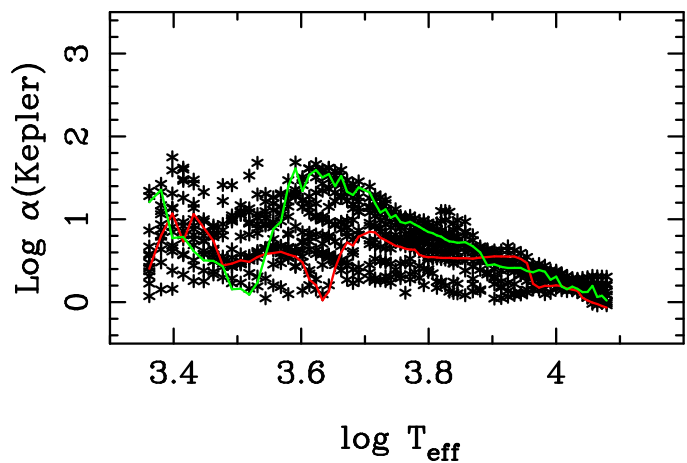}
        \caption{Same as in Fig.~5, but for the \textit{Kepler} passband.}
\end{figure}

\begin{figure}
        \includegraphics[height=8.9cm,width=6cm,angle=-90]{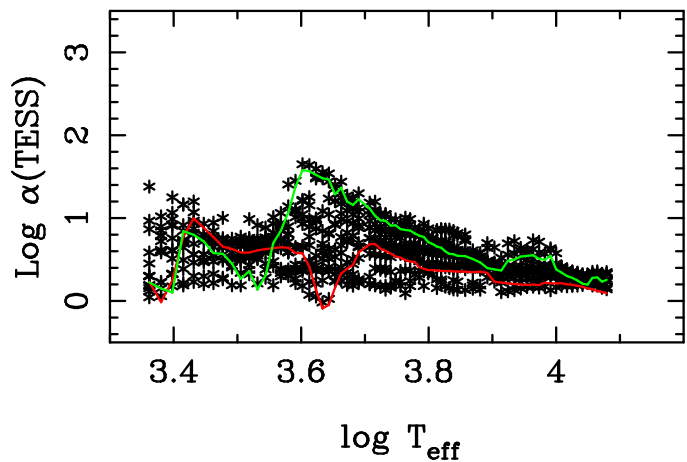}
        \caption{Same as in Fig.~5, but for the TESS passband.}
\end{figure}

\begin{figure}
        \includegraphics[height=8.9cm,width=6cm,angle=-90]{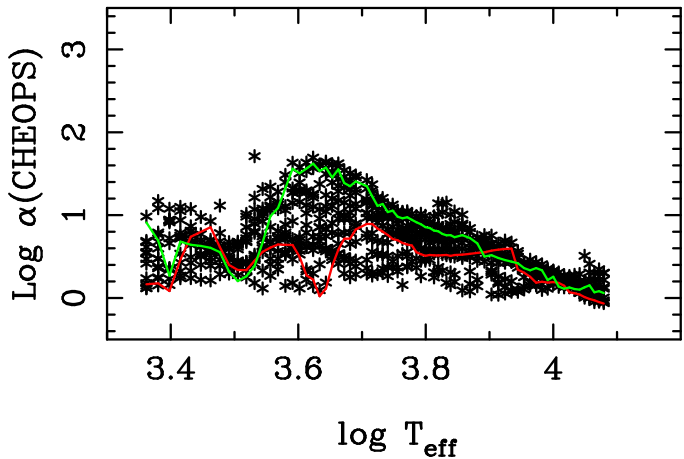}
        \caption{Same as in Fig.~5, but for the CHEOPS passband.}
\end{figure}

In Figs.\ 5--10, we trace the relative performance for two surface gravities: $\log g = 6.0$ for the most compact stars and $\log g = 3.5$ to represent subgiants. This shows how the relative quality of the fits depends on the surface gravity of a star. In general, the power-2 law is better at higher surface gravities, but is still not as good as the four-parameter law. This is the same in all passbands except for the interval $3.4 \gtrapprox \log T \gtrapprox 3.55$ in the $G_{\rm BP}$ band.  For models in radiative equilibrium, there is also a tendency for the relative quality of the fits to
not depend strongly on the surface gravity.

%%%%%%%%%%%%%%%%%%%%%%%%%%%%%%%%%%%%%%%%%%%%%%%%%%%%%%%%%%%%%%%%%%%%%%%%%%%%%%%%%%%%%%%%%%%%%%%%%%%%%%%%%%%%%%%%%%%%%%%%%%%%%%%%%%%%%%%%%%%%%%%%%%%%%%%%%%%%%%%%%%%%%%%%%%%%%%%%%%%%%%%%%%%%%%%%%%%%%%%%%%%%%%%%%%

\section{Comparison between plane-parallel and spherical LDCs}

A comparison between the power-2 LDCs predicted by plane-parallel and spherical model atmospheres is useful. For the plane-parallel model atmospheres, we adopted the results from Paper~I, which were calculated using the {\sc atlas} code. For the spherical model atmospheres, we used the results from the current work for both cases M1 and M2.

\begin{figure}
\includegraphics[width=\columnwidth]{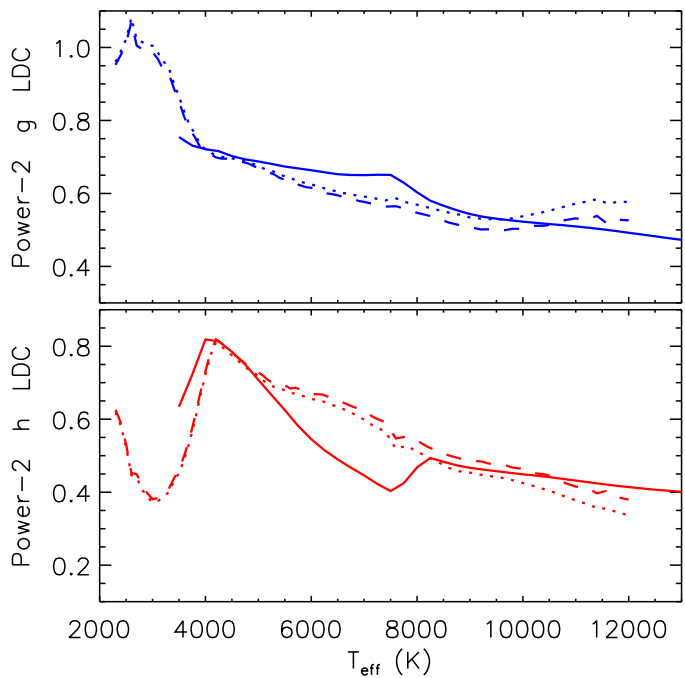}
\caption{\label{fig:gh} Comparison between the power-2 $g$ (upper panel) and $h$ (lower panel)
LDCs in the case of plane-parallel model atmospheres (solid lines) and spherical model atmospheres
for cases M1 (dotted lines) and M2 (dashed lines).}
\end{figure}

Figure~\ref{fig:gh} shows the variation of the $g$ and $h$ LDCs as a function of $T_{\rm eff}$, for $\log g=4.0$, solar metallicity, and $V_\xi = 2.0$\,km\,s$^{-1}$. Cases M1 and M2 show much closer agreement than the {\sc atlas} LDCs, in particular at lower $T_{\rm eff}$ values where spherical model atmospheres are more reliable. The anti-correlation between $g$ and $h$ is also clear, particularly at low $T_{\rm eff}$ values for M1 and M2, and at $6000 < T_{\rm eff} < 8000$~K for the {\sc atlas} LDCs.

\begin{figure}
\includegraphics[width=\columnwidth]{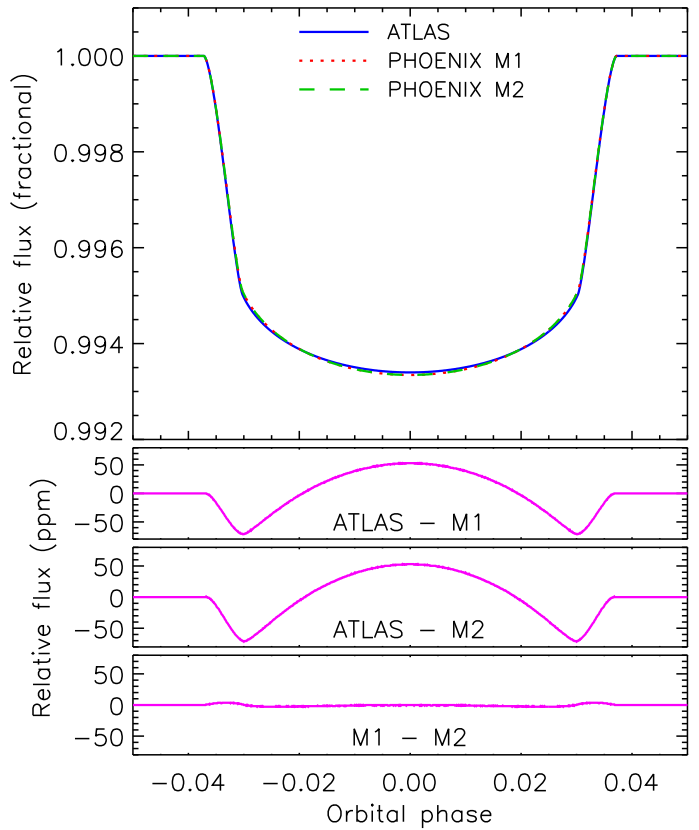}
\caption{\label{fig:JKTEBOP} Top panel: Comparison between the transit shapes using the power-2
LDCs for the plane-parallel ({\sc atlas}) and spherical (cases M1 and M2) model atmospheres
(labelled). Lower panels: Differences between the transit shapes in each case (labelled).}
\end{figure}

An important use of the LDCs presented in the current work is in calculating the light curves of planetary transits (see Fig.~\ref{fig:JKTEBOP}). We explored this by using the {\sc jktebop} code (Southworth 2013) to calculate the transit light curve of a system similar to HAT-P-7 (P\'al et al.\ 2008), using the transit parameters from Southworth (2011). This system was chosen because the small  ratio of the radii of the planet and star gives a good spatial sampling of the limb-darkening over the disc of the star. To avoid interpolating LDCs to specific atmospheric parameters, we adopted  fixed values of $T_{\rm eff} = 6000$\,K, $\log g=4.0$, solar metallicity, and $V_\xi = 2.0$\,km\,s$^{-1}$. The differences between the {\sc atlas} and {\sc phoenix} LDCs cause a change in the transit shape that reaches approximately 70\,ppm around second and third contact. In principle, this change is large enough to be measured from space-based missions such as {\it Kepler} and TESS, especially when many transits can be averaged together, but correlations between the measured LDCs and other parameters of the system will reduce the significance of the signal. The M1 and M2 LDCs give much more similar transit light curves ---the greatest deviation is only 4\,ppm---, which suggests that the choice of which to use is unimportant.

%%%%%%%%%%%%%%%%%%%%%%%%%%%%%%%%%%%%%%%%%%%%%%%%%%%%%%%%%%%%%%%%%%%%%%%%%%%%%%%%%%%%%%%%%%%%%%%%%%%%%%%%%%%%%%%%%%%%%%%%%%%%%%%%%%%%%%%%%%%%%%%%%%%%%%%%%%%%%%%%%%%%%%%%%%%%%%%%%%%%%%%%%%%%%%%%%%%%%%%%%%%%%%%%%%

\section{Summary and final remarks}

We computed LDCs for the power-2 law using specific intensities from spherical model atmospheres, as such results were only previously available for a small range of stellar parameters. To this end, we used the {\sc phoenix-cond} models for $T_{\rm eff}$ values from 2300\,K to 12000\,K, $\log g$ values from 0.0 to 6.0, solar metallicity, $V_\xi = 2.0$\,km\,s$^{-1}$, and with 900 $\mu$ points. We also computed LDCs for the four-parameter law, which supersede our previous calculations based on only 100 $\mu$ points. For both laws, we computed coefficients for the $uvby$, $UBVRIJHK,$ and SDSS $ugriz$ passbands, and for the \textit{Gaia}, \textit{Kepler}, TESS, and CHEOPS photometric systems. Two approaches were used to define the limb of the star based on the drop-off of intensity in the spherical model atmosphere predictions. We emphasise that the $\mu$ values have been rescaled to remove those beyond the limb of the star, and so this need not be done by users of the LDCs.

We performed comparisons between the merit function $\chi^2$ for the power-2 and the four-parameter laws, finding that the four-parameter law is superior in all cases to the power-2 law. The performance of the power-2 law is better at higher temperatures and at redder wavelengths. Additional calculations of LDCs for other photometric systems not covered in this paper are available upon request to the first author.

We recommend that the four-parameter law be used whenever high precision is needed for specific intensity profiles. If a less complex law is preferred, we recommend the power-2 law as it reproduces the theoretical intensity profiles to a higher precision than the other two-parameter laws (see Paper~I); we also recommend not using the linear and quadratic laws because they provide the least good fit to theoretical specific intensity profiles. For modelling eclipsing binary star and transiting planetary systems, the choice of which limb-darkening law to use has a relatively small effect on the best-fitting parameter values (Southworth 2023).

The current work, together with Paper~I, provides three sets of LDCs for any given set of stellar parameters and passbands, within those covered in this work. These are LDCs from plane-parallel {\sc atlas} model atmospheres (Paper~I) and LDCs from spherical model atmospheres (this work) calculated using methods M1 and M2. Those from methods M1 and M2 are very similar, whereas the LDCs from {\sc atlas} are moderately different. We show that these LDCs could be tested using high-precision photometry of transits or eclipses. In our experience, planetary transits are the best approach for this because the small size of the planet relative to the star provides a finer spatial sampling of the star's specific intensity profile, but the size of the differences between the transit light curves from different LDCs is not large.

%%%%%%%%%%%%%%%%%%%%%%%%%%%%%%%%%%%%%%%%%%%%%%%%%%%%%%%%%%%%%%%%%%%%%%%%%%%%%%%%%%%%%%%%%%%%%%%%%%%%%%%%%%%%%%%%%%%%%%%%%%%%%%%%%%%%%%%%%%%%%%%%%%%%%%%%%%%%%%%%%%%%%%%%%%%%%%%%%%%%%%%%%%%%%%%%%%%%%%%%%%%%%%%%%%

{}

%%%%%%%%%%%%%%%%%%%%%%%%%%%%%%%%%%%%%%%%%%%%%%%%%%%%%%%%%%%%%%%%%%%%%%%%%%%%%%%%%%%%%%%%%%%%%%%%%%%%%%%%%%%%%%%%%%%%%%%%%%%%%%%%%%%%%%%%%%%%%%%%%%%%%%%%%%%%%%%%%%%%%%%%%%%%%%%%%%%%%%%%%%%%%%%%%%%%%%%%%%%%%%%%%%

\begin{acknowledgements}
 We thank the anonymous referee for his/her helpful comments
that have improved the manuscript.
The Spanish MEC (AYA2015-71718-R, ESP2017-87676-C5-2-R, PID2019-107061GB-C64, and PID2019-109522GB-C52) is gratefully acknowledged for its support during the development of this work. A.C.\ acknowledges financial support from the grant CEX2021-001131-S funded by MCIN/AEI/ 10.13039/501100011033. This research has made use of the SIMBAD database, operated at the CDS, Strasbourg, France, of NASA's Astrophysics Data System Abstract Service and of SVO Filter Profile supported from the Spanish MINECO through grant AYA2017-84089.
\end{acknowledgements}

%%%%%%%%%%%%%%%%%%%%%%%%%%%%%%%%%%%%%%%%%%%%%%%%%%%%%%%%%%%%%%%%%%%%%%%%%%%%%%%%%%%%%%%%%%%%%%%%%%%%%%%%%%%%%%%%%%%%%%%%%%%%%%%%%%%%%%%%%%%%%%%%%%%%%%%%%%%%%%%%%%%%%%%%%%%%%%%%%%%%%%%%%%%%%%%%%%%%%%%%%%%%%%%%%%
\newpage
\newpage
\captionsetup[table]{labelformat=empty}
{Appendix A: Description of Tables 1-12 (available at the CDS)}
\begin{table*}
\renewcommand{\tablename}{Table A.}
        \caption{Power-2 and four-parameter LDCs for the \textit{Gaia}, \textit{Kepler}, TESS, CHEOPS, $ugriz$, $uvby,$ and $UBVRIJHK$ photometric systems.}
        \begin{flushright}
        \begin{tabular}{ccccccccc}
\hline
\scriptsize 
Name & Source & $T_{\rm eff}$ values (K) & log $g$ (c.g.s.) & [M/H] (dex) & $V_\xi$ (km\,s$^{-1}$) & Photometric system & Method/Eq.   \\
\hline
Table1 & {\sc PHOENIX} & 2300\,--\,12000 & 0.0\,--\,6.0 & 0.0 &2&{\textit{Gaia}, \textit{Kepler}, TESS, CHEOPS}& M1/{Eq. 2} \\
Table2 & {\sc PHOENIX} & 2300\,--\,12000 & 0.0\,--\,6.0 & 0.0 &2&{\textit{Gaia}, \textit{Kepler}, TESS, CHEOPS}& M2/{Eq. 2} \\
Table3 & {\sc PHOENIX} & 2300\,--\,12000 & 0.0\,--\,6.0 & 0.0 &2&{\textit{Gaia}, \textit{Kepler}, TESS, CHEOPS}& M1/{Eq. 3} \\
Table4 & {\sc PHOENIX} & 2300\,--\,12000 & 0.0\,--\,6.0 & 0.0 &2&{\textit{Gaia},  \textit{Kepler}, TESS, CHEOPS}& M2/{Eq. 3} \\

Table5 & {\sc PHOENIX} & 2300\,--\,12000 & 0.0\,--\,6.0 & 0.0 &2&{\textit{ugriz}}& M1/{Eq. 2} \\
Table6 & {\sc PHOENIX} & 2300\,--\,12000 & 0.0\,--\,6.0 & 0.0 &2 &{\textit{ugriz}}& M2/{Eq. 2} \\
Table7 & {\sc PHOENIX} & 2300\,--\,12000 & 0.0\,--\,6.0 & 0.0 &2 &{\textit{ugriz}}& M1/{Eq. 3} \\
Table8 & {\sc PHOENIX} & 2300\,--\,12000 & 0.0\,--\,6.0 & 0.0 &2 &{\textit{ugriz}}& M2/{Eq. 3} \\

Table9  & {\sc PHOENIX} & 2300\,--\,12000 & 0.0\,--\,6.0 & 0.0 &2 &{\textit{uvbyUBVRIJHK}}& M1/{Eq. 2} \\
Table10 & {\sc PHOENIX} & 2300\,--\,12000 & 0.0\,--\,6.0 & 0.0 &2 &{\textit{uvbyUBVRIJHK}}& M2/{Eq. 2} \\
Table11 & {\sc PHOENIX} & 2300\,--\,12000 & 0.0\,--\,6.0 & 0.0 &2 &{\textit{uvbyUBVRIJHK}}& M1/{Eq. 3} \\
Table12 & {\sc PHOENIX} & 2300\,--\,12000 & 0.0\,--\,6.0 & 0.0 &2 &{\textit{uvbyUBVRIJHK}}& M2/{Eq. 3} \\
\hline
\end{tabular}
\end{flushright}
\end{table*}

%%%%%%%%%%%%%%%%%%%%%%%%%%%%%%%%%%%%%%%%%%%%%%%%%%%%%%%%%%%%%%%%%%%%%%%%%%%%%%%%%%%%%%%%%%%%%%%%%%%%%%%%%%%%%%%%%%%%%%%%%%%%%%%%%%%%%%%%%%%%%%%%%%%%%%%%%%%%%%%%%%%%%%%%%%%%%%%%%%%%%%%%%%%%%%%%%%%%%%%%%%%%%%%%%%
\end{document}